\documentclass[twocolumn, times]{aastex631}

\newcommand{\mum}{\ifmmode{\rm \mu m}\else{$\mu$m}\fi}

\newcommand{\sevenrm}{\rm\scriptsize}

\newcommand{\NeV}{[Ne{\sevenrm\,V}]}
\newcommand{\NeII}{[Ne{\sevenrm\,II}]}
\newcommand{\NeIII}{[Ne{\sevenrm\,III}]}
\newcommand{\SIII}{[S{\sevenrm\,III}]}
\newcommand{\OIV}{[O{\sevenrm\,IV}]}
\newcommand{\OIII}{[O{\sevenrm\,III}]}

\received{Aug 1, 2022}
\revised{Sep 30, 2022}
\accepted{Oct 14, 2022}

\usepackage{CJKutf8}

\submitjournal{ApJL}

\shorttitle{Quasar Dust Emission from Narrow Line Regions}
\shortauthors{Lyu \& Rieke}

\begin{document}

\title{\large Polar Dust Emission in Quasar IR SEDs and Its Correlation with Narrow Line Regions}

\author[0000-0002-6221-1829]{Jianwei Lyu (\begin{CJK}{UTF8}{gbsn}吕建伟\end{CJK})}
\affiliation{ Steward Observatory, University of Arizona,
933 North Cherry Avenue, Tucson, AZ 85721, USA}

\author[0000-0003-2303-6519]{George H. Rieke}
\affiliation{ Steward Observatory, University of Arizona,
933 North Cherry Avenue, Tucson, AZ 85721, USA}

\begin{abstract}
Polar dust has been found to play an important role in the mid-infrared
emission of nearby Seyfert nuclei. If and how often polar dust exists among the
quasar population is unknown due to the lack of spatially-resolved
observations. In this Letter, we report correlations between the prominence of
AGN forbidden line emission (commonly associated with the narrow line region)
and the dust mid-IR energy output among the archetypal Palomar-Green quasar
sample and other bright type-1 AGNs drawn from the SDSS, Spitzer and WISE
archives. The AGN mid-IR color differences traced by WISE W2
($\sim4.6~\mum$)$-$W3 ($\sim12~\mum$) and W2 ($\sim4.6~\mum$)$-$W4
($\sim22~\mum$), and near-IR to mid-IR SEDs constrained with 2MASS, WISE and
Spitzer data have clear trends with the relative strength of the forbidden line
regions traced by the optical \OIII~and mid-IR \OIV~emission lines. These
observations indicate that, where the lines are strong, a large fraction of the
AGN emission at $\lambda\gtrsim5~\mu$m comes from dust in the forbidden line
regions. We find that the  widely
quoted universal AGN template is a result of averaging quasar SEDs with
different levels of polar dust emission above the torus output and that the typical 
intrinsic IR SED of compact torus dust emission alone falls with increasing wavelength 
past 5 $\mu$m (in $\nu F_\nu$). In addition, the association of polar dust with the 
forbidden lines suggests an alternative to the receding torus hypothesis for the decrease 
in infrared output with increasing AGN luminosity.
\end{abstract}



\section{Introduction}

In the unification model of active galactic nuclei (AGNs), the mid-IR emission
is assumed to originate from a compact torus-like obscuring structure
surrounding the central engine \citep{antonucci1993,urry1995}. This concept has
been widely adopted to interpret the infrared emission of AGNs (e.g.,
\citealt{fritz2006, nenkova2008,stalevski2012}; see reviews in
e.g., \citealt{netzer2015}). Ground-based high resolution imaging
of nearby Seyfert nuclei allows direct constraints on the dust morphology of
AGN mid-IR emission at $\sim$10~$\mu$m { down to a few tens of parsecs (pc),
and interferometry can reach pc-scale features}.  However, instead of a compact
{ pc-scale torus}, such observations have frequently revealed dust {
structures extending 10-100 pc} along the polar direction
\citep[e.g.,][]{bock2000,radomski2003,honig2013, lopez2016, asmus2016} and
challenged the classical picture.

In fact, the existence of AGN-heated dust on larger scales in the narrow line
region (NLR), preferentially along the polar direction, was suggested long ago
\citep[e.g.,][]{netzer1993, dopita2002} { and directly revealed by, e.g.,
    the early spatially-resolved mid-IR observations of the archetypal type-2
    AGN NGC 1068 and type-1 AGN NGC 4151 where the extended dust emission at
    $\sim$10~$\mum$ is co-spatial with the NLR morphology
    \citep[e.g.,][]{bock2000, radomski2003}. Recent observations of larger AGN
    samples frequently report that the extended mid-IR emission is aligned with
    NLR position angle, suggesting this behavior is characteristic of many AGNs
    \citep[e.g.,][]{lopez2016, asmus2016, alonso2021}.  Notably,
    \cite{asmus2019} selected AGNs with strong mid-IR \OIV$\lambda$25.89~$\mu$m
    emission for mid-IR interferometry observations and found polar emission in
    nearly every member of the sample, supporting the  association of AGN
mid-IR polar emission with NLRs.}

Nevertheless,  addressing the prevalence of extended polar dust is difficult
observationally, as mid-IR observations with sufficient spatial resolution are
limited to the nearest objects with preferred inclination angles. { However,
    \citet{lyu2018} suggest that the polar dust component can leave distinct
    features in the AGN IR spectral energy distribution (SED) to allow its
identification, and argue that the typical NLR clouds can
naturally host the dust grains responsible for the polar dust emission.} If
confirmed, such an association could extend our understanding of the incidence
of polar dust to much more distant objects { and allow new insights into the
AGN IR SED variations}.

In this Letter, we investigate whether there is a direct connection between the
dust responsible for the AGN mid-IR emission and the gas producing the narrow
line emission.  \S~\ref{sec:method} summarizes the various optical and mid-IR
measurements of AGNs and quasars used for our analysis. We report a correlation
between the relative strengths of the AGN NLRs and the behavior of the mid-IR
output in \S~\ref{sec:result}. \S~\ref{sec:discussion} discusses how this
result reshapes our understanding of various AGN properties in general.

\section{Sample, Data and Measurements}\label{sec:method}

Our study is focused on three samples:
\begin{itemize}
    \item PG quasars: this sample consists of the 87 $z<0.5$ Palomar-Green (PG)
        quasars selected from the Palomar Bright Quasar Survey
        \citep{schmidt1983, boroson1992}, which have rich IR photometric data
        from 2MASS, UKIDSS, {\it Spitzer}, {\it Herschel} and {\it WISE} and
        high-quality {\it Spitzer}/IRS spectra in the mid-IR. These sources
        have served  as a foundation for quasar studies over the past four
        decades. In \cite{lyu2017, lyu2017b}, we used this dataset to study the
        intrinsic AGN IR SED variations of quasars in detail,  laying a basis
        for the current work;

    \item SDSS quasars: To increase the sample statistics, we also select 8506
        quasars from the DR7 edition of the Sloan Digital Sky Survey (SDSS)
        Quasar Catalog, with  spectroscopy compiled by  \cite{shen2011}. We
        require these objects to be at $z<0.5$ and to have measurements of
        \OIII~and H$\beta$. We have identified near- to mid-IR counterparts
        within a search radius of 3$''$ in 2MASS and WISE for 7712 of these
        quasars (90.7\%);

    \item SDSS-Spitzer AGNs: We also build a third type-1 AGN sample
        with mid-IR spectra by matching the Spitzer/IRS archive with the
        broad-line AGNs from the SDSS survey. These type-1 AGNs are identified
        by FWHM($H\alpha$)$~\gtrsim1200$~km/s, either in the DR7 edition of the
        SDSS Main Galaxy Sample \citep{hao2005} or the Quasar Catalog
        \citep{shen2011}. The mid-IR spectra are mainly from the Combined Atlas
        of Sources with Spitzer IRS Spectra (CASSIS;
        \citealt{cassis2011})\footnote{\url{https://cassis.sirtf.com/}} with
        some additional objects from the ATLAS project
        \citep{hernan2011}\footnote{\url{http://www.denebola.org/atlas/}}.  We
        have visually inspected the mid-IR spectra to make sure that they have
        adequate signal-to-noise ratios for the $\sim$6--30 $\mu$m continuum as
        well as the mid-IR \OIV~emission line, and do not have overwhelmingly
        strong PAH emission or silicate absorption features. In total, there
        are 209 type-1 AGNs under these criteria.
\end{itemize}

To trace the NLR, we use forbidden lines of oxygen, \OIII $\lambda$5007\AA~and
\OIV$\lambda$25.89$\mu$m; both lines have relatively low ionization energy and
extend over large  physical scales ($\gtrsim100$~pc) around AGNs.  As both
lines { show correllations with the AGN luminosity}
\citep[e.g.,][]{diamond2009, malkan2017}, basing the study on absolute line
strengths is not appropriate.  Instead, we normalize each of them with {
another AGN} luminosity indicator, namely H$\beta$  and the 4--5 $\mu$m
continuum (traced by the WISE W2 band)\footnote{The strength of the former is
    dominated in type-1 AGNs by the emission of broad-line regions and the
    latter probes the hot dust emission strength near the AGN sublimation zone,
    both of which { trace common nuclear structures in type-1 AGNs below or
        near parsec scales \citep[e.g.,][]{kaspi2005, lyu2019} and are believed
to be approximate indicators of AGN luminosity}\citep[e.g.,][]{malkan2017,
asmus2015}.}, to remove this dependence { and trace the relative strength of
the NLR}. 

The optical emission line measurements for PG quasars are taken from
\cite{boroson1992} and for the SDSS quasars from \cite{shen2011}. The mid-IR
\OIV~ properties { of the sample}, such as line flux, FWHM and equivalent
width (EW), are measured by fitting a single Gaussian function with a local
continuum to the observed Spitzer/IRS data with our own customized code.

More detailed information on the sample as well as the relevant data and measurements
are provided online at \url{https://github.com/karlan/AGN-MIR-SED-NLR.
}

\section{Results}\label{sec:result}

To characterize the AGN IR emission, we first calculate the IR colors for
individual sources and later compute average SED/spectral templates for objects
binned by different NLR strengths.

\subsection{Trend of Infrared Colors with NLR Strength}

To trace the significance of polar infrared emission, we adopt the WISE W2$-$W3
and W2$-$W4 color differences, since (1) W3 ($\sim12\mum$) probes the
wavelength range where the existence of polar dust is well established and (2)
W4 ($\sim22\mum$) is near the peak of the expected polar dust SED
\citep[e.g.,][]{honig2013, lyu2018}.

\begin{figure}[htp]
    \begin{center}
  \includegraphics[width=1.0\hsize]{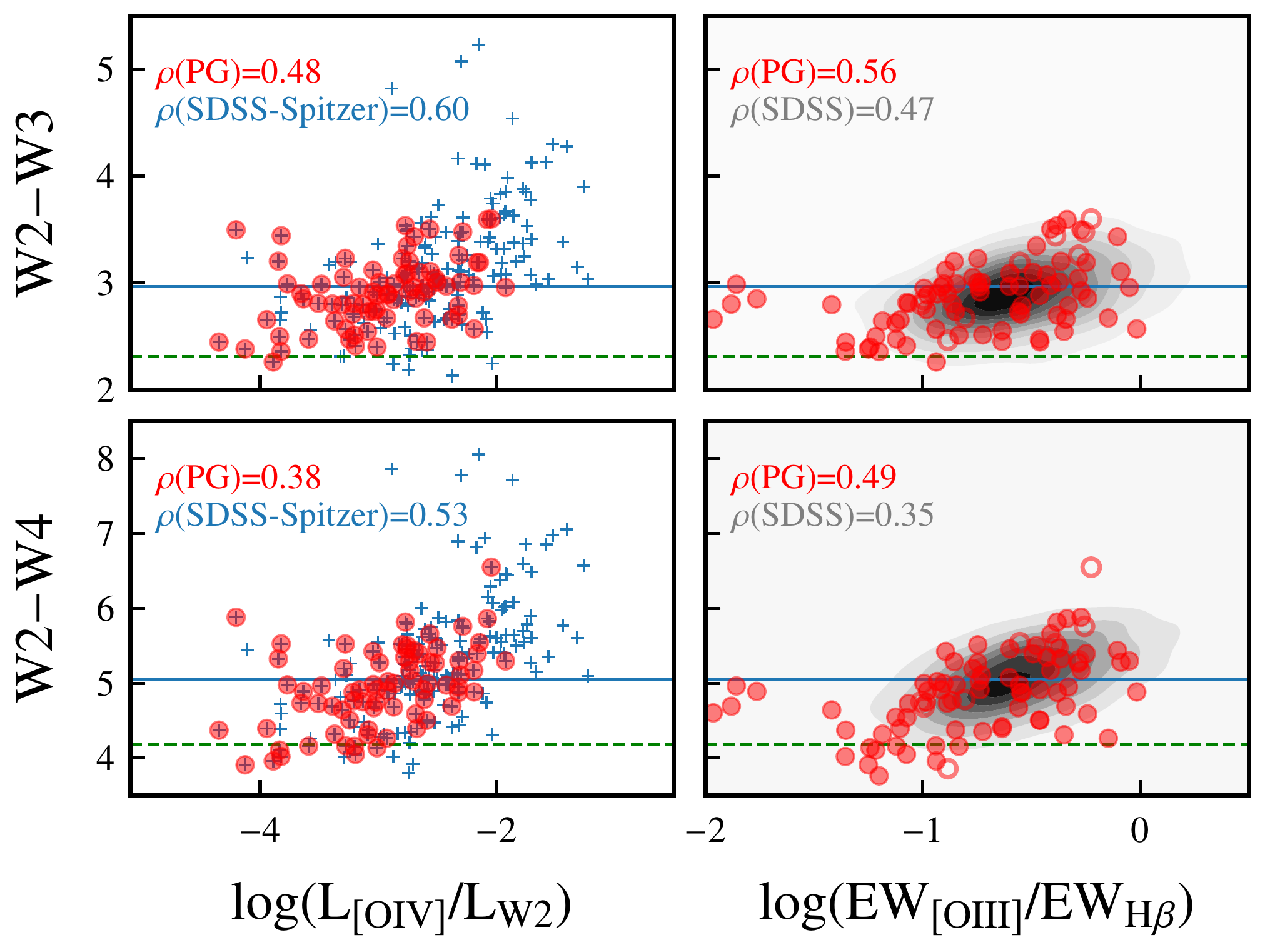}
    \caption{WISE colors vs. \OIV 25.89$\mu$m/W2 luminosity ratio (left panels)
        and \OIII 5007\AA/H$\beta$ EW ratio (right panels). The red dots are
        for the PG quasar sample, with the open circles indicating the cases  where galaxy
        emission contributes over $50\%$ of the IR output (from \cite{lyu2017}),
        the blue crosses are for the SDSS-Spitzer sample (left panels) and
        background grey contours are the distribution of the SDSS quasar sample
        (right panels). { The typical 1-$\sigma$ uncertainties of these measurements
        are $\lesssim$ 5--10\%. } The green-dashed and blue solid lines are,
        respectively, the template colors of warm-dust-deficient and normal AGN
        templates at the average redshift of the sample. The Spearman's rank
        correlation coefficients $\rho$ of each sample are denoted in the
    corresponding color.}
  \label{fig:quasar-o34}
    \end{center}
\end{figure}

The left panels in Figure~\ref{fig:quasar-o34} show how the WISE W2$-$W3 and
W2$-$W4 colors indicate stronger mid-IR emission with increasing  \OIV/W2  for
both the PG sample and the SDSS-Spitzer sample. Correlations are apparent in
both panels. The Spearman's rank correlation coefficients $\rho$ in the W2$-$W3
bands for the PG and SDSS-Spitzer quasars are 0.476 ($p$=1.383e-05) and 0.601
($p$=6.03e-22), where $p$ is the probability of the null hypothesis that a
correlation does not exist. The Spearman's $\rho$ in W2$-$W4 for PG and
SDSS-Spitzer quasars are 0.382 ($p$=6e-4) and 0.531 ($p$=1.282e-16).

Similar AGN IR color trends also exist for the optical NLR tracer,
EW(\OIII)/EW(H$\beta$), for both PG and SDSS quasars, as presented in the right
panels of Figure~\ref{fig:quasar-o34}. The Spearman's $\rho$ values for the
W2$-$W3 color difference  for PG and SDSS quasars are 0.487 ($p$=2.98e-06) and
0.349 ($p$=1.1e-239), respectively. The values for W2$-$W4 for PG and SDSS
quasars are 0.561 ($p$=0.0) and 0.471 ($p$=3.459e-08), respectively. For the PG
sample, we have highlighted those cases with a strong host contribution in the
IR with open circles, as measured in \citep{lyu2017b}. For quasars with $f_{\rm
host, IR}>0.5$, the AGN still dominates the mid-IR by  $>$ 90\%
\citep{lyu2017b}. That is, the influence of galaxy contamination on these
trends is minimal.

\subsection{Average SED Templates}

To illustrate further how the forbidden line strength is connected with the nature of
the AGN SED, we sorted the PG and SDSS-Spitzer samples into bins of
$L_{\OIV}/L_{W2}$ and EW(\OIII)/EW(H$\beta$) and derived the corresponding AGN
templates for different relative NLR strengths. The vast majority of these
objects have {\it Spitzer}/IRS spectra that can reveal host galaxy
contamination in the mid-IR emission by the detection of the 11.3 $\mu$m PAH
feature. To remove the galaxy contamination, we subtracted a \cite{rieke2009}
star-forming galaxy IR template from the observed {\it Spitzer}/IRS spectra to
make the peak of any residual 11.3 $\mu$m PAH feature strength fall below
3-$\sigma$ flux uncertainities. As the {\it Spitzer} spectra only span the
rest-frame 5.5--35~$\mu$m, we extended the wavelength coverage at shorter
wavelengths with the photometric SEDs from 2MASS and WISE. We interpolated the
IR photometry of each object in $\log\nu-\log F_\nu$ space, smoothed the
interpolated SED with a $\Delta\log(\nu)=0.2$ boxcar with the same wavelength
grid and replaced it with the mid-IR spectra if available. 

We visually investigated these SEDs and dropped any objects that (1) show
abnormal SED jumps due to time variability (note that the 2MASS/WISE/Spitzer
    data are observed at different times; (2) have mid-IR silicate features in
    absorption; (3) have very strong PAH features that dominate the mid-IR
    emission; (4) show convincing evidence for non-thermal mid-IR emission
    (e.g., some BL Lac objects have a falling power-law mid-IR continuum); (5)
    do not have measurements of the mid-IR \OIV~line strength.  After
    incoporating all these contraints, 177 objects were left. We grouped them
    into three bins for $L_{\OIV}/L_{W2}$ or EW(\OIII)/EW(H$\beta$) with
    identical sample sizes ($\sim$59--60 objects), normalized each SED at
    rest-frame 14.5$\mu$m where the spectrum is featureless, and computed the
    median average SEDs as the final templates.

\begin{figure*}[htp]
    \begin{center}
        \includegraphics[width=1.0\hsize]{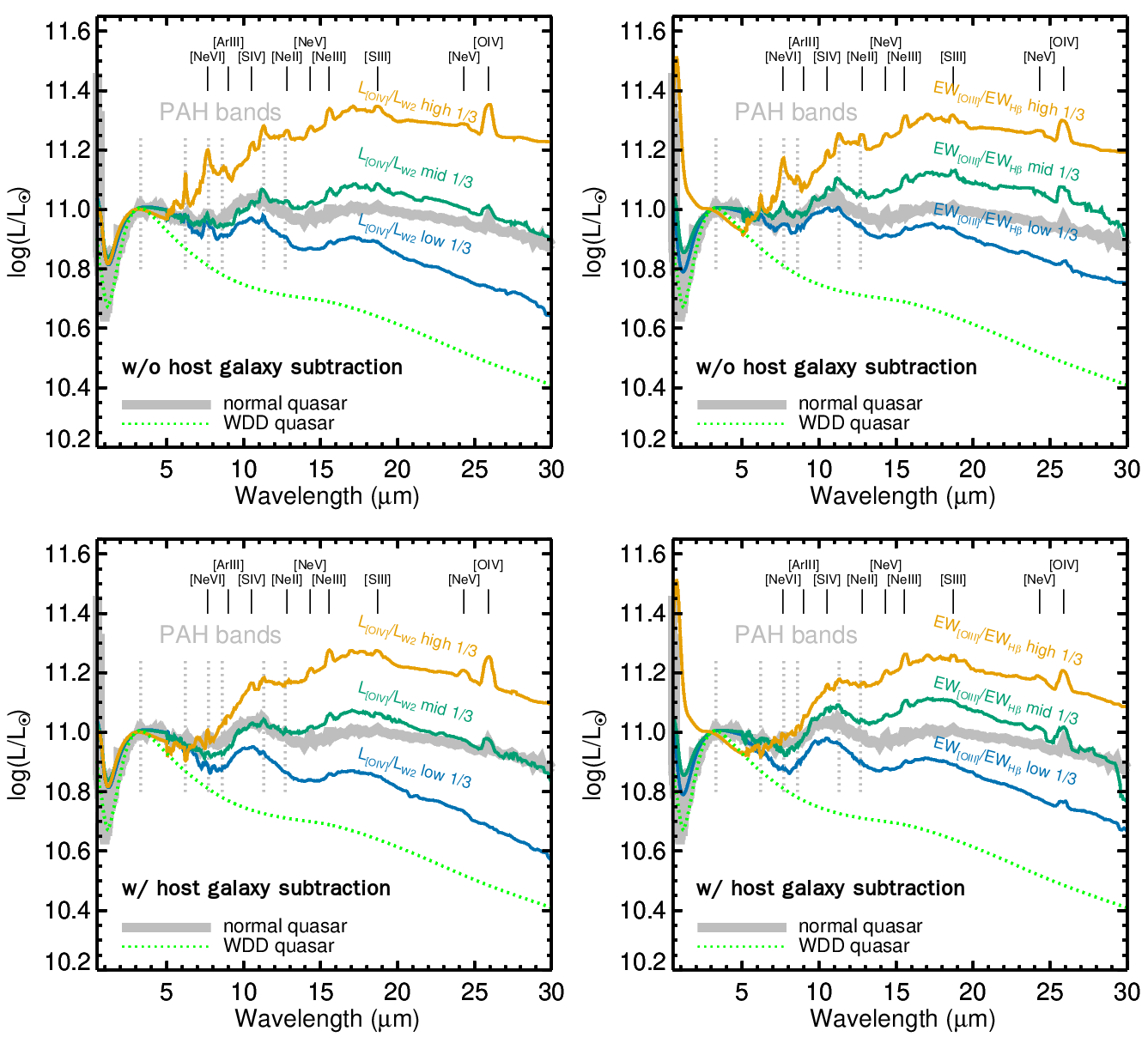}
    \caption{ Average infrared SEDs of the sample binned by
        $L_{\OIV}/L_{W2}$  (left panels) and of EW(\OIII)/EW(H$\beta$) (right
        panels) without (top panels) and with host galaxy subtraction (bottom panels). 
        The top 1/3, middle 1/3 and low 1/3 of the population are
        plotted as orange, dark green and blue solid lines.  As a comparision, we
        also show the spectral template of the normal quasar population (thick
        grey line) and the broad-band SED template of the warm-dust-deficient (WDD) quasar population (light green dotted line)
        constructed in \cite{lyu2017}.  All the templates are normalized at
        3~$\mum$ with common mid-IR emission features denoted.}
  \label{fig:quasar_sed_o4}
    \end{center}
\end{figure*}

{ In Figure~\ref{fig:quasar_sed_o4}, we show the average templates before
(upper panels) and after removing the host galaxy contamination (bottom
panels). Perceivable trends with $L_{\OIV}/L_{W2}$ and EW(\OIII)/EW(H$\beta$)
are present in all the panels. After the host galaxy removal,} the resulting
AGN SED templates have very weak PAH features and the continuum emission should
be dominated by the AGN-heated dust. We can see clearly that the relative
strength of the AGN mid-IR (10--30~$\mu$m) output to that at  $\sim$3~$\mum$
increases  with increasing \OIV/W2 ratio or \OIII/H$\beta$ ratio. The behavior
is consistent with the color trends in Figure~\ref{fig:quasar-o34}, but
attaches specific SEDs to the trend.  In addition,  the prominance of the
mid-IR forbidden lines, e.g., \NeII, \NeIII, \NeV~and \SIII, is also associated
with an increase of mid-IR continuum strength.

{
\subsection{Robustness of These Correlations}
\label{sec:robust}

The trends in Figure~\ref{fig:quasar-o34} and Figure~\ref{fig:quasar_sed_o4}
should not be affected by extinction. Our study is focused on Type-1 to 1.5
AGNs where the Balmer broad lines are clearly detected and the extinction in
the visible is relatively small. As the dust opacity decreases quickly from the
optical to the mid-IR (e.g., \citealt{Gordon2021}), the expected extinction in
the W2 and W4 bands and the [OIV] line would be even smaller and can be safely
ignored. The W3 band does include the silicate absorption feature but the
extinction over the entire broad band is still far less ($<$ 6\%; e.g.,
\citealt{rieke1985}) than in the visible. The [OIII]/H$\beta$ ratio is widely
used as being roughly extinction-independent due to their close wavelengths.

The behavior of EW(\OIII)/EW(H$\beta$) in Figure~\ref{fig:quasar-o34} might be
affected by the Baldwin Effect. As shown by e.g.,
\citet{kova2010}, the Baldwin Effect is similar for \OIII~and narrow H$\beta$.
The broad H$\beta$, however, has a weak and negative Baldwin Effect. The net
result would drive down \OIII/H$\beta$  modestly as the AGN luminosity
increases. In fact, within the errors the slopes of the ratio of line flux to
continuum flux are identical for these lines, i.e., the ratio is independent of
luminosity \citep{wilkes1999}.  The Baldwin Effect is less thoroughly observed
for the \OIV  ~ line, although \citet{kere2009} observe it for other
mid-infrared fine structure lines. Again, it will drive down the ratio of the
\OIV ~line to W2 with increasing luminosity.  However, in both cases since the
shapes of the infrared SEDs of AGNs , e.g., W2 $-$ W3 or W2 $-$ W4, do not
correlate with luminosity\footnote{Although the strength of the infrared SED
relative to the visible does correlate with luminosity, see
\S~\ref{sec:covering}.}, the effect will increase the scatter in
Figure~\ref{fig:quasar-o34} but should not contribute to the correlations.

Figure~\ref{fig:quasar_sample} compares the redshift and AGN luminosity
distributions of the sub-samples used to compute the average templates
in Figure~\ref{fig:quasar_sed_o4}. No statistical difference is seen for the sample redshifts. 
However, the AGN luminosity shows a negative
trend with both $L_{\OIV}/L_{W2}$ and EW(\OIII)/EW(H$\beta$). As discussed in
\S~\ref{sec:covering}, the NLR strength does have some dependence on the
AGN luminosity and thus this AGN luminosity difference is a result of
sub-sample definition. For the AGN mid-IR SED and NLR strength  correlations
presented above, however, we used the ratios of two luminosity indicators to
take out this luminosity effect. As a result, our interpretation is not
influenced by the sample luminosity.

\begin{figure*}[htp]
    \begin{center}
        \includegraphics[width=0.49\hsize]{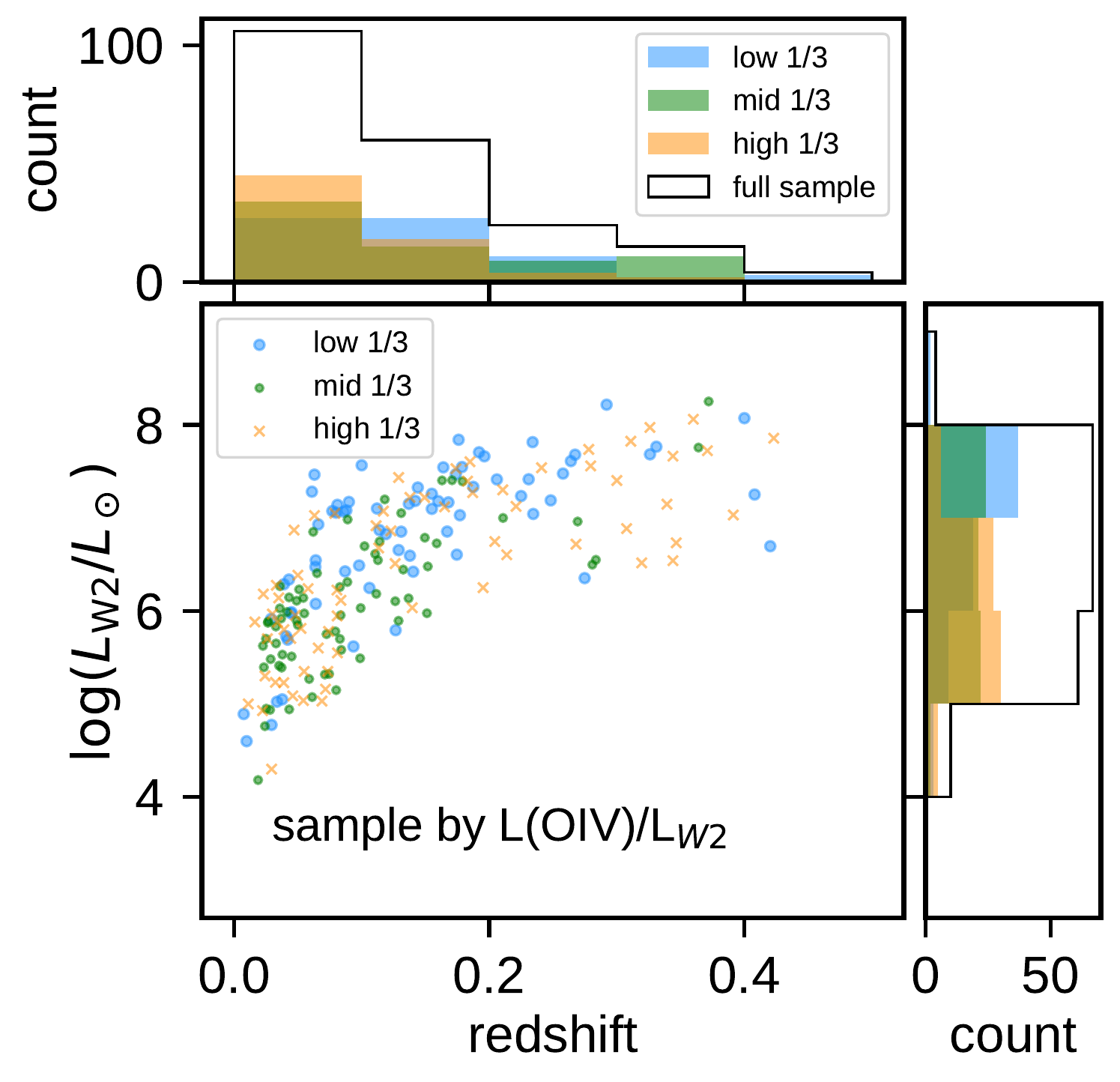}
        \includegraphics[width=0.49\hsize]{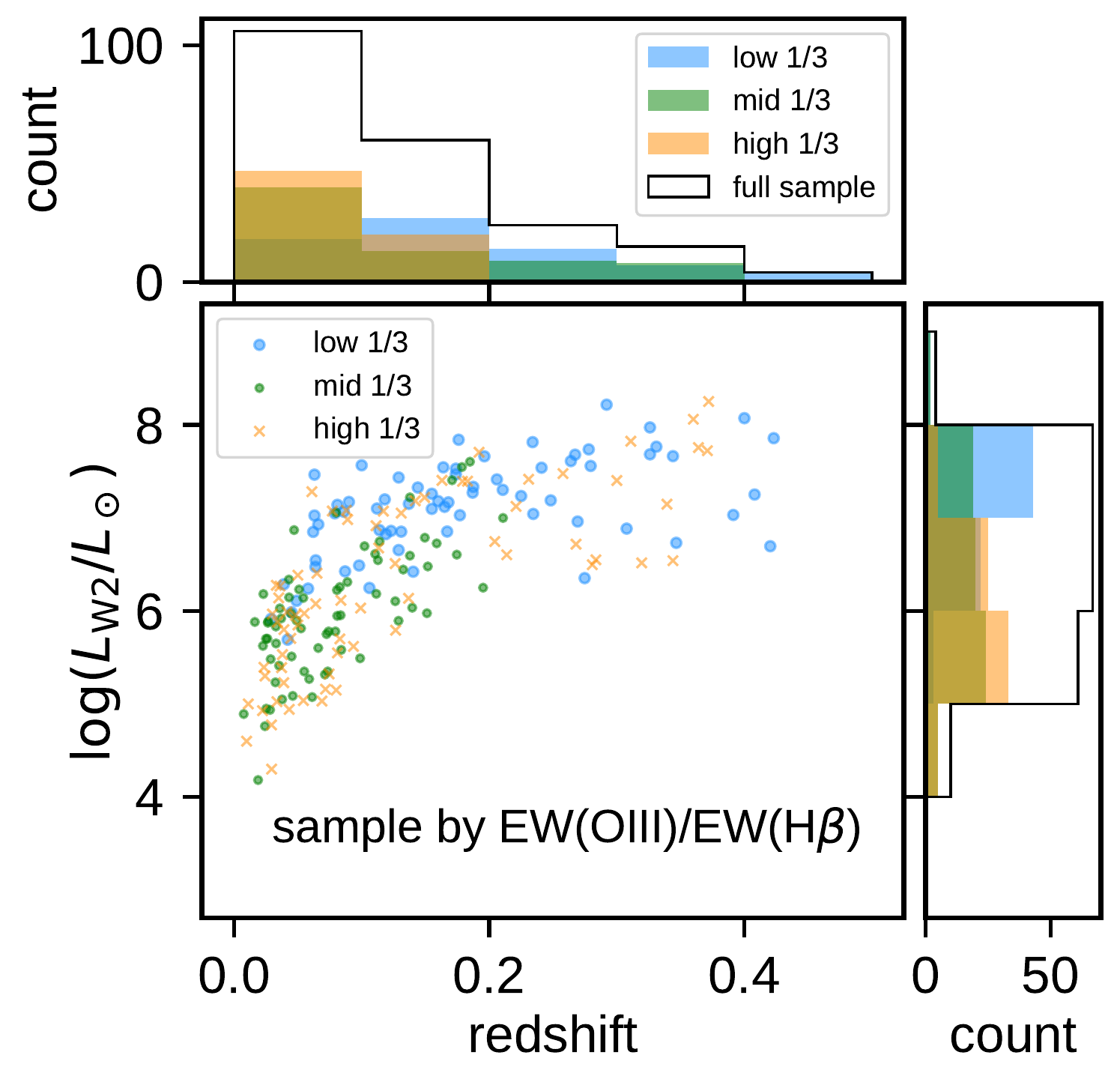}
    \caption{ Redshifts and luminosity (traced by $L_{\rm W2}$) distribution of the sub-samples used to 
        compute AGN templates in Figure~\ref{fig:quasar_sed_o4}.  
    }
  \label{fig:quasar_sample}
    \end{center}
\end{figure*}

Lastly, the choice of host galaxy template would not affect our results as the changes
of the AGN color after the galaxy subtraction are small, as shown in
Figure~\ref{fig:quasar-o34}.

In conclusion, none of the factors mentioned above can explain the
correlations.

}

\section{Discussion}\label{sec:discussion}

Previous work by \citet{zhang2013} reported a correlation between the covering
factor for the dust around an AGN and the strength of the wing of the
\OIII$\lambda$5007\AA~ line relative to the bolometric luminosity. The results
obtained in this work agree generally with that study and our analysis reveals
a strong correlation between the relative AGN mid-IR energy output and the
relative strength of the NLR emission; i.e., the more prominent the NLR region,
the stronger the dust emission in the mid-IR.  As shown below, this result
expands our understanding of the variations in AGN SEDs we have described
previously and can provide new insights to AGN behavior.

\subsection{Quasar IR SEDs}

In \cite{lyu2017}, we found two major variations in  intrinsic IR SEDs compared
with ``normal'' AGNs \citep[e.g., that of][]{elvis1994}:  (1) the
warm-dust-deficient (WDD) case has similar hot dust emission at 1--3~$\mum$ as
normal objects but is weaker in the mid-IR; and (2) the hot-dust-deficient
(HDD) case has weak dust emission throughout the near- and mid-IR. As shown in
Figure~\ref{fig:quasar-o34}, the WDD template represents the smallest W2$-$W3
and W2$-$W4 color differences seen in quasars; { it also } appears to be the
limiting case for a weak NLR. This suggests that the WDD template represents
the compact dust emission associated with the classical torus alone. In
\citet{lyu2018} we showed that the normal AGN SED can be approximately
reproduced by adding polar dust emission above the WDD template with an
effective optical depth $\tau_{\rm pol}\sim0.8$.  { Consistent with this
possibility,} our semi-empirical SED fitting for all three type-1 AGNs with
polar dust constraints from  mid-IR interferometry (NGC 3783, ESO 323-77 and
NGC 4507) suggested that their nuclear dust emission SED could be WDD. 

Particularly for the comparison based on the \OIV~ line, the SEDs for weak line
emission drop toward longer wavelengths in Figure~\ref{fig:quasar_sed_o4}
similarly to the behavior of the HDD and WDD templates. In comparison, the SED
for strong line emission lies above even the ``normal'' template of
\citet{lyu2017}.  Based on this evidence, we suggest that a large fraction, if
not all, the excess emission above the WDD template comes from { relatively}
extended dust associated with the NLR.

The average quasar SEDs derived in the literature show similar dust continua
that are described by the normal template and have little evolution with
redshift \citep[e.g., see references in][]{lyu2018}. Given the discussion in
the preceding paragraph, this suggests that a large fraction of quasars have a
moderate level of polar dust emission. Some 60--70\% of the PG quasars have
excess AGN dust emission above the WDD template and the average relative
fraction of this hypothetical  extended dust emission is $\sim$50\% at $\sim$
10~$\mum$, from the comparison of normal and WDD AGN templates \citep[see
Figure 2 in][]{lyu2017}.  As most works computing average quasar templates do
not differentiate SED properties, the normal-like AGN templates are a result of
averaging AGN SEDs with different levels of extended NLR dust emission and
different intrinsic torus SEDs, including cases that are relatively brighter in
the mid-infrared than even the normal template. This situation is illustrated
in \citet{shang2011}, Figure 10, and  \citet{lyu2018}, Figures 6 \& 7. The
success of the normal template in SED fitting and decomposition derives
directly from its being an overall average, not because it represents all AGNs
accurately.

A deeper understanding of the dusty surrounds of AGNs requires decoding the
roles of the circumnuclear tori and the polar dust. High resolution imaging
\citep[e.g.,][]{bock2000, radomski2003, alonso2021} and interferometry
\citep[e.g.,][]{honig2013,asmus2019,isbell2022} of nearby AGNs have established
that the mid-infrared emission of many of them is dominated by polar dust often
shown to be associated directly with their NLRs, rather than being solely the
output of compact ($\lesssim1-10$ pc) tori. Extending such studies to a
significant number of distant and luminous AGNs is not feasible, but the
association of mid-infrared color differences with the relative strength of the
forbidden oxygen lines shows that a similar situation holds for them. This
association is not predicted at all by traditional torus models.

\subsection{Role of the NLR and of Winds}

From a study of $\sim$4200 type-1 AGNs from SDSS, \cite{zhang2013} also
reported correlations between the mid-IR dust emission strength and optical
\OIII~emission, in their case both normalized by the AGN bolometric luminosity.
They decomposed the \OIII~emission profiles into core and wing components and
found that the correlations are driven by the wing component rather than the
core. However, we find similar or even higher correlation coefficients for the
entire \OIII~emission line, possibly because we  compare the W3 and W4 fluxes
with the W2 one, i.e, testing the shape of the mid-IR SED rather than the role
of the mid-IR in the total luminosity.

\cite{zhang2013} concluded that dusty outflows launched near the torus, not the
more extended NLR, contribute most of the AGN mid-IR emission. However,
observations of nearby AGNs such as NGC 1068 \citep{bock2000}, NGC 4151
\citep{radomski2003}, and many more \citep{alonso2021} show that a substantial
fraction of the 10 $\mu$m emission comes from regions well removed from the
nucleus, up to tens of pc for luminous systems.  This behavior is consistent
with our finding substantial correlations with the integrated oxygen line
strengths rather than just the broad components.

\citet{baron2019} also used the \OIII ~and WISE data to explore the role of
winds. Their sample of AGNs was matched closely in narrow \OIII ~luminosity
between cases with only narrow \OIII ~and those with a significant broad
component in the \OIII~ line profile, indicative of winds. They compare color
differences in the WISE bands for these two samples. There is a convincing case
for the systems with winds to have stronger emission in W4 (22 $\mu$m) than
those without but mixed results for a stronger excess in W3 (12 $\mu$m).

\citet{alonso2021} compare high resolution images at $\sim$ 10 $\mu$m with ALMA
images of submm emission by dust and the distribution of CO (3-2). The variety
of configurations and overall complexity undermines any universal and simple
interpretation. They find in the most extended cases that the 10 $\mu$m
emission has sizes of 50 $-$ 160 pc, much less than the kpc sizes derived from
the models of \citet{baron2019}. However, the groundbased data are insensitive
to low surface brightness extended source components.

The polar dust emission seen by mid-IR interferometry is on a different scale
from the 50 $-$ 160 pc dusty narrow line regions seen in imaging.  However, as
pointed out by \cite{lyu2018}, this extended dust structure is expected to have
a temperature gradient that makes the dust emission size change as a function
of the observed wavelength. In fact, a simple SED model featuring diffuse dust
distributed over 100 pc scales proposed in the same work matched the observed
polar dust sizes at $\sim10~\mum$.  This argument is illustrated by the GRAVITY
\citep{gravity2020}, MATISSE  \citep{gamez2022}, and imaging \citet{bock2000}
observations of NGC~1068. The first reveals a ring-like structure at K-band
($\sim2.2~\mu$m), the second shows dust emission at $\sim4$--12$\mum$ mostly
consistent with a polar wind, and the third shows extended emission on 100 pc
scales to the full extent of the bright forbidden line region seen at 12.5 $\mu$m
and suggests even greater extent at 24.5 $\mu$m. Further extent at low surface
brightness at 12.8 $\mu$m is also shown by \citet{galliano2005}. 

The questions about the interpretation of the extended mid-IR emission should
be answered by JWST/MIRI observations, which will obtain images of sufficient
resolution and surface brightness sensitivity out through 21 $\mu$m, thus
including the spectral range where, from the work of \citet{baron2019}, the
wind-associated emission should be brightest.

\subsection{Dust Covering Factor vs Luminosity}
\label{sec:covering}

From comparing AGN IR templates, we can estimate the dust covering factor of
the NLR for a typical quasar with a normal IR SED.  From the calculations in
\cite{lyu2017, lyu2017b}, the ratio of dust emission to the AGN accretion disk
luminosity $L_{\rm NLR dust, IR}/L_{\rm accr. disk}$  is 0.53 and 0.37 for the
normal and WDD templates, respectively.  Assuming the WDD template represents
the SED of the compact torus and any excess IR emission is from the NLR, this
would indicate about 37\% and 15\% of the accretion disk emission is obscured,
i.e. ``covered'', by the equatorial torus and the polar NLR dust, respectively.
In other words, the torus vs. polar dust emission luminosity is about 2.3:1 for
the normal quasar population.

Although there is little evidence for a strong redshift evolution of AGN IR SED
behavior, the relative strength of the overall AGN mid-IR emission is commonly
found to decrease with source luminosity \citep[e.g.,][]{maiolino2007,
roseboom2013}. In the past, a so-called receding torus has frequently been
invoked to explain this observation \citep[e.g.][]{lawrence1991}. It posits
that the torus dust is destroyed by the increasing AGN luminosity, resulting in
less IR emission.  Given the very likely connection between the AGN mid-IR
emission and dusty NLR, this behavior is likely to arise instead from the
well-established reduction in the NLR prominence with  increasing AGN
luminosity \citep[e.g.,][]{netzer2006, stern2012}.

\begin{figure}[htp]
    \begin{center}
  \includegraphics[width=1.0\hsize]{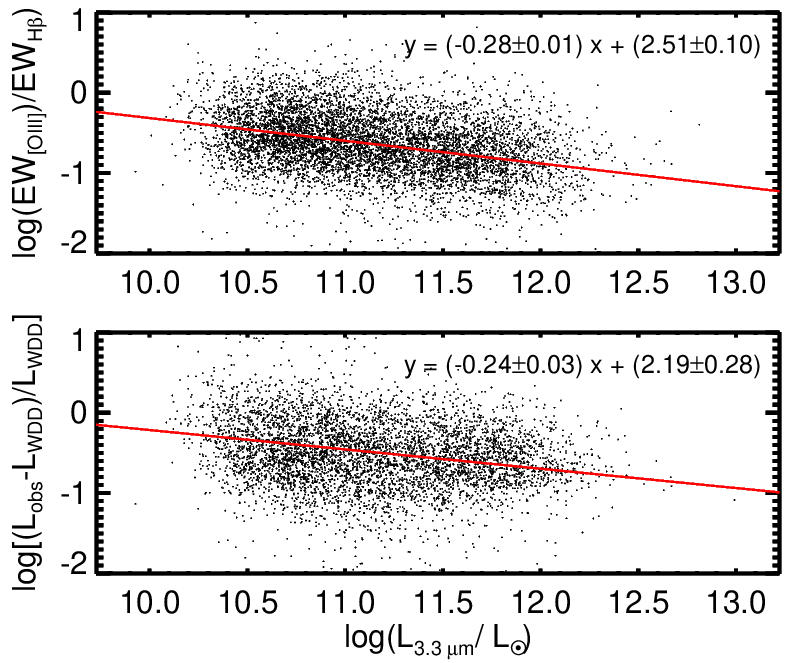}
    \caption{Luminosity dependence of the AGN narrow-line emission strength and
        IR excess emission. For each quasar, we have interpolated and smoothed
        the observed SEDs, subtracted a WDD AGN template with the normalization
        at rest-frame 3.3~$\mu$m, and computed the { relative excess}
    luminosity at 5--25~$\mu$m as $(L_{\rm obs}-L_{\rm WDD})/L_{\rm WDD}$.  The
red lines show the linear regressions of the data points with the fitted
relations reported in the top-right corner of each panel.}
  \label{fig:nlr_lum}
    \end{center}
\end{figure}

To illustrate this possibility, we present the luminosity dependence of the
\OIII~relative strength and { IR relative excess emission} above the WDD
template for the SDSS quasar sample in Figure~\ref{fig:nlr_lum}. Both
parameters show clear { negative trends with AGN luminosity with Spearman's
    rank correlation coefficient values $\rho$=$-$0.42 ($p<$1e-4) and
    $\rho$=$-$0.21 ($p<$1e-4), respectively. { The Baldwin Effect on
    \OIII/H$\beta$ is inadequate to account for any significant part of the
behavior in the upper panel of the figure (see \S~\ref{sec:robust}).}
Fitting these data by linear regression, we find the slopes of the correlations
are consistent within uncertainties.} In conjunction with the correlations
between these two properties shown in Figure~\ref{fig:quasar-o34}, this
behavior also provides further evidence for the mid-IR emission of lower
luminosity AGNs having a substantial component from their NLRs. 

\section{Summary}\label{sec:summary}

We have demonstrated a correlation between the relative strengths of the
optical \OIII~and mid-IR \OIV~lines and the AGN mid-IR color or SED shape. This
indicates that, on average, AGNs with strong forbidden line emission also have
a large fraction of emission at $\lambda>5~\mum$ emitted by dust in the
forbidden line regions. The AGNs with weak forbidden lines have infrared SEDs
that drop past 5 $\mu$m with increasing wavelength in $\nu F_\nu$, compared
with the flat or rising SEDs on average for those with strong forbidden lines.
This difference suggests that the compact circumnuclear torus emits a SED
similar to the weak forbidden line case, i.e., as in the warm dust deficient
(WDD) template of \citet{lyu2017}.  It also implies that $\sim$60--70\% of the
optically-selected blue quasars have a significant level of polar dust emission
with a relative contribution at $\sim10~\mum$ about 50\% on average above the
WDD template. For a normal quasar with a typical average SED, the fraction of
accretion disk UV-optical emission transferred into the infrared is estimated
to be 37\% and 15\% by the equatorial torus and polar dusty NLR, respectively.

We also find that the decrease of the so-called dust covering factor with the
increase of AGN luminosity is likely to be a result of the reduction of the
forbidden  line region rather than a receding torus. 

We expect upcoming JWST/MIRI spatially-resolved observations of nearby AGNs to
greatly advance our understanding of the extended dust structures surrounding
AGNs.

\begin{acknowledgments}
    We thank for anonymous referee for the comments and suggestions. This work
    was supported by NASA grants NNX13AD82G and 1255094.
\end{acknowledgments}

\eject

\end{document}